\documentclass[12pt]{article}
\usepackage{graphicx}

\begin{document}
\font\little=cmr10

\title{TIME-FRACTIONAL DIFFUSION \\ OF DISTRIBUTED ORDER}
\vskip -2.0truecm

\author{
Francesco MAINARDI\thanks{Corresponding author,
E-mail: {\tt francesco.mainardi@unibo.it}},
\ Antonio MURA,
\\ {\little Department of Physics, University of Bologna, and INFN,}
\\ {\little Via Irnerio 46, I-40126 Bologna, Italy}
\and
Gianni PAGNINI
\\ {\little Italian Agency for  New Technologies,
  Energy and the Environment} 
\\ {\little (ENEA), Via Martiri di Monte Sole 4,
 I-40129 Bologna, Italy}
 \\
 \and
Rudolf GORENFLO
\\ {\little Department of  Mathematics and Informatics, Free University of  Berlin,} 
\\  {\little Arnimallee  3, D-14195 Berlin, Germany}
}
\vskip -1.5truecm

\date{December 2006 - September 2007}

\maketitle
\vskip -0.8truecm

\centerline{\little International Workshop on  Fractional  Differentiation and its Applications}
\centerline{\little (FDA06), 19-21 July 2006, Porto, Portugal.}
\centerline{{\bf Journal of Vibration and Control, in press (2007).}}
\vskip -2.0truecm 

\begin{abstract}
\noindent
{\little The partial differential equation of 
Gaussian diffusion  is generalized by using
 the time-fractional derivative
 of distributed order
between 0 and 1, in both  the Riemann-Liouville (R-L) and the Caputo (C)  sense. 
For a general distribution
of time orders we  provide
 the fundamental solution,  
 that is still a probability density, in terms of an integral
 of Laplace type. 
  The  kernel   depends on the type of the assumed fractional derivative
 except for the single order case where the two  approaches turn to be equivalent.       
We  consider with some detail  two cases 
of order distribution: the double-order   and
the uniformly distributed order.
For these cases 
we exhibit plots of the corresponding fundamental solutions
and their variance, pointing out the remarkable difference between  the two approaches
for small and large times.} 

\end{abstract}


\def\pni{\par \noindent}
\def\vsh{\smallskip}
\def\vs{\medskip}
\def\vvs{\bigskip}
\def\vvvs{\bigskip\medskip} 
\def\vsp{\vsh \pni}
\def\vsn{\vsh\pni}
\def\cen{\centerline}
\def\ra{\item{a)\ }} \def\rb{\item{b)\ }}   \def\rc{\item{c)\ }}
\def\q{\quad} \def\qq{\qquad}
\def \rec#1{{1\over{#1}}}
\def\ds{\displaystyle}
\def\eg{{\it e.g.}\ }
\def\ie{{\it i.e.}\ }
\def\versus{{\it vs.}\ }
\def\e{{\rm e}}
\def\d{\partial}
\def\dx{\partial x}    \def\dt{\partial t}
\def\Ai{{\rm Ai}\,}
\def\Erfc{{\rm Erfc}\,}
 \def\E{{\rm E}}
 \def\EE{{\mathcal E}}
\def\Ei{{\rm Ei}\,}
\def\Ein{{\rm Ein}\,}
\def\log{{\rm log}\,}
\def\u{\widetilde{u}}
\def\ul{\widetilde{u}} 
\def\uf{\widehat{u}} 
\def\r{\right} \def\l{\left}
\def\rt{\right} \def\lt{\left}
\def\lra{\Longleftrightarrow}
\def\RR{\vbox {\hbox to 8.9pt {I\hskip-2.1pt R\hfil}}}
\def\NN{{\rm I\hskip-2pt N}}
\def\CC{{\rm C\hskip-4.8pt \vrule height 6pt width 12000sp\hskip 5pt}}
\def\L{{\mathcal L}} 
\def\F{{\mathcal F}}  
\def\M{{\mathcal M}}  


\font\bfs=cmbx12 scaled\magstep1

\font\note=cmr10 at 10 truept  
\font\note=cmr8  

\newpage

\section{Introduction}

It is well known that the fundamental solution (or {\it Green function})
for the {Cauchy} problem of the   linear diffusion equation
can be interpreted as a Gaussian (normal)  probability density function ($pdf$)
in space, evolving in time.
All the moments of this $pdf$ are finite;
in particular, its variance
is proportional to the first power of time, a noteworthy property
of the {\it standard diffusion}.
\vsp
In this  paper we illustrate via fractional calculus
two types of generalization  of this Cauchy problem.
One type    uses the fractional derivative in the sense of Riemann and
    Liouville (R-L), the other in the sense of   Caputo (C). In its uses we
    distinguish between single and distributed orders of fractional
    derivatives.  
Specifically, we  work out how to express
their fundamental solutions in terms
of an integral of Laplace type  suitable for a numerical
evaluation.
Particular attention is devoted to  the time evolution
of  the variance for the R-L and C cases. It is known  that for large times
the variance  characterizes the type of anomalous diffusion.
	\vsp
The plan of the paper is as follows. 
\vsp
In Section 2 we write down the two  general forms of the time-fractional
diffusion equation with distributed-order with R-L and C derivatives
and the Fourier-Laplace representation of the corresponding  fundamental solution. 
For this purpose  we need to introduce a positive function $p(\beta)$
that acts as a discrete or continuous distribution of orders.
In addition to the particular case of a single order $\beta _0$ with
$0<\beta _0\le 1$,
 we consider two case-studies for the fractional diffusion of distributed order:
as a discrete distribution we take two distinct  orders $\beta_1, \beta_2$
with $0<\beta_1< \beta_2 \le 1$;
as continuous distribution we take the uniform density 
of orders between zero and 1.  
\vsp
Section 3 is devoted to the time evolution   of the variance which is obtained
 from the Fourier-Laplace representation of the corresponding  fundamental solution,
 by inverting only the Laplace transform.
 In the single order case we recover the sub-diffusion power-law  common
 to the R-L and C forms;  for the cases of distributed order
 we find a remarkable difference between the two forms, well visible
 from their asymptotic expressions for  small and large times.  
In section 4 we illustrate our method to get the fundamental solutions
 from  their Fourier-Laplace transforms, following the strategy
 of carrying out at first the Fourier inversion and then the Laplace inversion. 
We find instructive to show  the graphical representation of the fundamental
solutions (in space at fixed times).
For the case of fractional diffusion  of single order,
because of the self-similarity of the solutions we limit ourselves  to
show plots of the corresponding solutions  at a fixed time $t=1$ versus $x$.
For the two case-studies of fractional relaxation of distributed order,
because the self-similarity of the solutions is lost, 
we provide plots of the corresponding solutions  versus $x$ at three fixed times, selected
as $t=0.1$, $t=1$ and $t=10$, contrasting  the different evolution for the R-L
and    the C form, in a  moderate space-range.
  We can note how the time evolution
 of the solution  in the considered spatial range depends  on the different 
 time-asymptotic behaviour of the variance  for the two forms.
\vsp
Finally, concluding remarks are given in Section 5.
\vsp	 
	In order to have a self-contained mathematical treatment   
we have added three Appendices: 
the Appendix A is
devoted to the basic notions  of fractional calculus,
whereas Appendices  B and C deal with  functions of Mittag-Leffler
and Exponential Integral type, respectively,
 in view of their  relevance  for our treatment.


\section{The equations for time-fractional diffusion  of distributed order}

\subsection{The R-L and the C  forms in space-time domain}
 The {\it standard  diffusion equation}, that
in  re-scaled non-dimensional variables reads 
$$ \frac{\d}{\dt}\, u(x,t)
   =   {\d^2  \over \dx^2 } \,u(x,t)\,, \;
 x \in \RR \,,\; t \in \RR_0^+\,,
    \eqno(2.1)$$
        with $u(x,t)$ as  the field variable,
can be generalized by using
the notion of  fractional derivative of distributed order
in time\footnote{
We find an earlier idea of fractional derivative of distributed order
in time in the 1969 book by Caputo \cite{Caputo 69},
that was later developed by Caputo himself,
see  \cite{Caputo FERRARA95,Caputo FCAA01}
and by Bagley \& Torvik, see \cite{BagleyTorvik 00}
A basic framework for the numerical solution
of distributed-order differential equations has been recently introduced
by Diethelm \& Ford  \cite{Diethelm-Ford FCAA01},
Diethelm \& Luchko \cite{Diethelm-Luchko 04}
andby Hartley \& Lorenzo [ \cite{Hartley-Lorenzo SP03,Lorenzo-Hartley NLD02}.
}. 
\vsp
For this purpose we need to consider a function $p(\beta)$
that acts as weight for the order of differentiation  $\beta \in (0,1]$
such that
$$  p(\beta)\ge 0\,, \q \hbox{and} \q
   \int _0^1 \! p(\beta )\,d\beta =  c >0\,. \eqno(2.2)$$
The positive constant $c$ can be taken as 1
if we like to assume the normalization condition for the integral.
 Clearly, some special conditions of regularity and behaviour near
 the boundaries will be required for the weight function
   $p(\beta)$\footnote{
   For the weight function $p(\beta)$
   we    conveniently  require    that its primitive
$ P(\beta) = \int_0^\beta \!\!   p(\beta^\prime)\, d \beta^\prime $
vanishes at $\beta=0$ and is there continuous from the right,
attains the value $c$ at $\beta=1$
and has at most finitely many  (upwards) jump points in 
the half-open interval $0<\beta\le 1$,
these jump points allowing delta contributions to 
$p(\beta)$ 
(particularly relevant for discrete distributions of orders).}.
Such function, that can be referred to as the {\it order density} if $c=1$,
 is allowed to have $\delta$-components
if we are interested in  a discrete distribution of orders.
\vsp
There are two possible forms of generalization depending if we
use  fractional derivatives  intended in the R-L or C sense. Correspondingly
 we  obtain
the {\it  time-fractional diffusion equation of distributed order}
in the  two forms:
$$
  \frac{\d}{\dt} \, u(x,t)
    =
\int_0^1 p(\beta)\,_tD^{1-\beta}
\left [\,\frac{\d^2}{\d x^2}\,u(x,t) \right ]\,d\beta
   \,, \q   x\in \RR, \; t\ge 0\,,
\eqno(2.3a) $$
and
$$
  \int _0^1 p(\beta )\,\left [\, _tD_*^\beta  \, u_*(x,t)\right ] \, d\beta
\,    =
\frac{\d^2  }{\d x^2 }   \,u_*(x,t)\,, \q  x\in \RR, \; t\ge 0\,.
\eqno(2.3b) $$
From now on we shall restrict our attention on the
fundamental solutions of Eqs. (2.3a)-(2.3b) so
we understand that these equations are
subjected to the initial condition $ u(x,0^+) = 
 u_*(x,0^+) = \delta  (x)$.
 Since for distributed order the solution depends
on the selected form (as we shall show hereafter),
we now  distinguish the two fractional equations
and their fundamental solutions
by decorating in the Caputo case
the variable $u(x,t)$ with subscript $*$
as it is customary for the notation of the corresponding derivative.  
\vsp
Diffusion equations of  distributed order
of both types have been recently discussed by several authors:
in particular  we find the C form e.g.
in \cite{Caputo FCAA01,ChechkinGorenfloSokolov PRE02,ChechkinGorenfloSokolovGonchar FCAA03,%
ChechkinKlafterSokolov EUROPHYSICS03,Naber FRACTALS04,SokolovChechkinKlafter POL04}
whereas
the R-L form e.g.
in \cite{Langlands PhysA06,SokolovChechkinKlafter POL04,SokolovKlafter CHAOS05}.
In some papers  the authors have referred  to the C and R-L forms 
as to {\it normal} and {\it modified}
forms of the time-fractional diffusion equation
of distributed order, respectively.   
\vsp
For a  thorough   general study of
fractional pseudo-differential equations of distributed order
let us cite the  paper
by Umarov and Gorenflo \cite{Umarov-Gorenflo ZAA05}.
For a   relationship with the Continuous Random Walk models
we may refer to the  paper by Gorenflo and Mainardi
\cite{GorenfloMainardi CARRY05}.

\subsection{The  RL and C forms in Fourier-Laplace domain}

The  fundamental solutions for  the time-fractional diffusion equations
(2.3a)-(2.3b) can be obtained
by applying in sequence
the Fourier and Laplace transforms to  them.
We  write, for generic functions $v(x)$ and $w(t)$,
these transforms as follows:
$$
\begin{array}{ll}
&  { \ds \F \left\{ v(x);\kappa \right\} = \widehat v(\kappa)
  := \int_{-\infty}^{+\infty}\!\! \e^{\,\ds i\kappa x}\,v(x)\, dx\,,
  \;\kappa \in \RR\,,} \\  \\
&  {\ds \L \left\{ w(t);s \right\} = \widetilde w(s)
  := \int_{0}^{+\infty} \!\!\e^{\,\ds -st}\,w(t)\, dt\,,
  \;s \in \CC\,.}
\end{array}
$$
Then,
in the Fourier-Laplace domain our  Cauchy problems
[with $u(x,0^+) = u_*(x,0^+)= \delta (x)$],
after applying   formulas for the
Laplace transform  appropriate  to the R-L and C fractional derivatives,
see (A.8') and (A.9), 
and observing $\widehat \delta (\kappa ) \equiv 1$, see e.g. \cite{Gelfand-Shilov 64},
appear in the two forms
$$  s \widehat{\widetilde{u}}(\kappa ,s) -1 = -\kappa^2 \,
\left[\int _0^\infty p(\beta) s^{1-\beta}   \, d\beta \right]
\, \widehat{\widetilde{u}}(\kappa ,s) \,,\eqno(2.4a)$$ 
$$  \left[\int _0^\infty p(\beta) s^\beta   \, d\beta\right] \, \widehat{\widetilde{u}}(\kappa ,s)
 -   \int _0^\infty p(\beta) s^{\beta -1}\,d\beta
    = -\kappa^2\,
   \widehat{\widetilde{u}}(\kappa ,s) \,. \eqno(2.4b)
$$
Then, introducing the relevant functions
$$ A(s) = \int _0^1 \! p(\beta )\, s^{1-\beta} \,d\beta  \,, \eqno(2.5a)$$
and
$$ B(s) =  \int _0^1 \! p(\beta )\, s^\beta \,d\beta  \,, \eqno(2.5b)$$
we then get for the R-L and C cases 
the Fourier-Laplace representation of the corresponding fundamental solutions:
$$  \widehat{\widetilde u}(\kappa,s) =
\frac{ 1}{s + \kappa^2  A(s)}=  \frac{1/A(s)}{\kappa^2 + s/A(s)}\,,
\eqno(2.6a)$$
and
$$  \widehat{\widetilde u}_*(\kappa,s) =
\frac{ B(s)/s}{\kappa^2 + B(s)}\,.
\eqno(2.6b)$$
From Eqs. (2.6a)-(2.6b) we recognize that the passage between the R-L and  the  C form
can be carried out  by the transformation
$$  \left\{ \hbox{C}: \;B(s)\right\}    \, \Longleftrightarrow \, 
     \left\{\hbox{R-L}:\; \frac{s}{A(s)} \right\} \,.\eqno (2.7)$$ 
We note that in the particular case of time fractional diffusion 
of single order $\beta_0$ ($0<\beta _0\le 1$) we have $p(\beta)= \delta(\beta -\beta _0)$
hence  in (2.5a): $A(s)= s^{1-\beta _0}$, in (2.5b): $B(s)= s^{\beta _0}$,
so that $ B(s) \equiv s/A(s)$.
Then,  Eqs. (2.6a) and (2.6b) provide the same  result
$$ \widehat{\widetilde u}(\kappa,s) \equiv  \widehat{\widetilde u}_*(\kappa,s) =
\frac{ s^{\beta _0 -1}}{\kappa^2 + s^\beta_0}\,.
\eqno(2.8)$$
This is consistent with the well-known result according to
which the two forms are equivalent for the single order case.
However, for a generic order distribution, the Fourier-Laplace representations
(2.6a) (2.6b) are different so they produce in the space-time domain
different fundamental solutions, that however are interrelated in some way 
in view of the transformation (2.7).


\section{The variance of the fundamental solutions}
\subsection{General considerations}
Before  trying to get the fundamental solutions in the space-time domain
to be obtained by a double inversion of the Fourier-Laplace transforms,
 it is worth to outline the expressions of their  second moment 
 (the variance)
since these  can be derived from Eqs. (2.6a)-(2.6b) by a single Laplace inversion,
as it is shown hereafter. 
We recall that the time evolution of the variance  is relevant    
for classifying the type of  diffusion. 
\vsp
Denoting  for the two forms 
$$  \hbox{R-L}\,:\; \sigma^2(t):= \int_{-\infty}^{+\infty}\!\! x^2\, u(x,t)\,dx \,,
\q \hbox{C}\,: \;
\sigma ^2_*(t):= \int_{-\infty}^{+\infty}\!\! x^2\, u_*(x,t)\,dx\,,
\eqno(3.1)$$
we easily recognize that
$$   \hbox{R-L}\,: \; \sigma^2 (t)=
   -{\ds  \frac{\d^2}{\d \kappa^2}\, \widehat{u}(\kappa=0,t)}\,, 
  \q  \hbox{C}\,: \;
 \sigma _* ^2 (t)=
   -{\ds  \frac{\d^2}{\d \kappa^2}\,
 \widehat{u}_*(\kappa=0,t)}\,. 
 \eqno(3.2)$$
As a consequence we  need to invert only Laplace transforms 
taking into account the behaviour of the Fourier transform for $\kappa $ near zero.
 \\
 For \underbar{the R-L case} we get from Eq. (2.6a),
 $$ \widehat{\widetilde{u}}(\kappa,s) =
\frac{1}{s} \, \left (1 - \kappa^2\,\frac{A(s)}{s} + \dots \right )\,,$$
so we obtain
$$ \widetilde{\sigma ^2}(s) =
- \frac{\d^2}{\d \kappa^2} \widehat{\widetilde{u}}(\kappa=0,s)=
  \frac{2A(s)}{s^2}\,.\eqno(3.3a)$$
\\
For \underbar{the C-case} we get from Eq. (2.6b)
$$ \widehat{\widetilde{u}}_*(\kappa,s) =
\frac{1}{s} \, \left (1 - \kappa^2 \,\frac{1}{B(s)} + \dots \right )\,,$$
so we obtain
$$
\widetilde{\sigma^2}_*(s) =
- \frac{\d^2}{\d \kappa^2} \widehat{\widetilde{u_*}}(\kappa=0,s)=
  \frac{2}{s\, B(s)}.\eqno(3.3b)$$
Except  for the single order diffusion, were we  recover the well-know result
 $$\sigma ^2 (t) \equiv  \sigma  _* ^2(t)  = 2\, \frac{t^{\, \ds \beta _0} }{\Gamma(\beta _0 +1)}\,,
\q 0<\beta _0 \le 1\,, \eqno(3.4)$$
for a generic order distribution, we expect that the time evolution
of the variance substantially  depends on the chosen (R-L or C) form.   
\vsp
We shall now concentrate our interest
to some typical choices for the  weight function $p(\beta )$
 that characterizes 
 the time-fractional diffusion  equations  of distributed order
(2.3a) and (2.3b).  
This will allow us to compare the results for the
R-L form and for the C form.   

\subsection{Fractional diffusion  of double-order}

First, we  consider the choice
$$
p(\beta ) =p_1\delta (\beta -\beta_1 )+ p_2\delta (\beta -\beta_2)\,, \q
0<\beta _1<\beta _2\le 1\,, 
 \eqno(3.5)$$
where the constants $p_1$ and $p_2$ are both positive,
conveniently  restricted to the normalization condition $p_1+p_2=1$.
\vsp
Then for \underbar{the R-L case} we have
$$
A(s) = p_1 \,s^{1-\beta _1} + p_2\, s^{1-\beta _2}\,, \eqno(3.6a)
$$
so, in virtue of (3.3a),
 we have 
 $$
\widetilde {\sigma^2}(s) = 2\,  p_1 \,s^{-(1+\beta _1)} + 2\, p_2\, s^{-(1+\beta _2)}\,. \eqno(3.7a)$$
Finally the Laplace inversion yields,
see and compare 
 \cite{Langlands PhysA06,SokolovChechkinKlafter POL04},
  $$
 \sigma^2(t) = 2\,  p_1 \,\frac{t^{\beta _1}} {\Gamma(\beta_1+1)} + 
 2 \,p_2\, \frac{t^{\beta _2}} {\Gamma(\beta_2+1)}
  \sim  
\cases{
{\ds 2 p_1\, {\ds\frac{ t^{\beta _1}}{\Gamma(1+\beta _1)}}}\,,&
 $ \!\! t \to 0^+\,,$ \cr
{\ds 2 p_2\,\frac{t^{\beta _2}}{\Gamma(1+\beta _2)} }\,,&
 $ \! \! t \to +\infty\,.$ 
}
  \eqno(3.8a)$$  
Similarly, for \underbar{the C case} we have
$$
B(s) = p_1 \, s^{\beta _1} + p_2 \,s^{\beta _2} \,,
\eqno(3.6b)
$$
so, in virtue of (3.3b),
$$
\widetilde {\sigma^2}_*(s) = \frac{2}{  p_1 \,s^{(1+\beta _1)} + p_2\, s^{(1+\beta _2)}}
\,. \eqno(3.7b)$$
Finally the Laplace inversion yields,
 see and compare \cite{ChechkinGorenfloSokolov PRE02},  
 $$
\sigma^2_*(t) = {\ds \frac{2}{  p_2 }\,t^{\beta _2}}\, 
 {\ds E_{\beta_2-\beta_1, \beta_2+1}\lt( -\frac{p_1}{p_2}\, t^{\beta _2-\beta_1}\rt)} 
  \sim  
\cases{
{\ds \frac{2}{p_2}\, \frac {t^{\beta _2}}{\Gamma(1+\beta _2)}}\,,&
 $\!\! t \to 0^+\,,$\cr
{\ds \frac{2}{p_1}\,\frac{t^{\beta _1}}{\Gamma(1+\beta _1)}}\,,&
 $\!\! t \to +\infty\,.$ 
}
  \eqno(3.8b)$$
Then we see that for the R-L case we have  an explicit combination of two power laws:
the smallest exponent($\beta_1$) dominates at small times whereas the largest exponent
($\beta_2$) dominates at large times.
For the C case we have a Mittag-Leffler function in two parameters
so we have a combination of two power laws only asymptotically for small and large times;
precisely we get a behaviour   opposite to the previous one,
so   the largest exponent($\beta_2$) dominates at small times whereas the smallest exponent
($\beta_1$) dominates at large times.  
 \vsp
We can derive the above asymptotic behaviours directly from the Laplace transforms
(3.7a)-(3.7b) by applying the Tauberian theory
for Laplace transforms\footnote{
According to this theory
 the asymptotic behaviour of a function $f(t)$ near $t=\infty$
and $t =0$
is (formally)  obtained from the asymptotic behaviour
of its Laplace transform $\widetilde f(s)$ for $s \to 0^+$ and
for $s \to +\infty$, respectively.} . 
In fact for  the R-L case 
 we note  that for $A(s)$ in (3.6a)  ${\ds s^{1-\beta_1}}$ is  negligibly
small in comparison with  ${\ds s^{1-\beta_2}}$for $s \to 0^+$ and, viceversa,
   ${\ds s^{1-\beta_2}}$ is  negligibly small
in comparison to ${\ds s^{1-\beta_1}}$for $s \to +\infty$.
Similarly for the C case
we note  that for $B(s)$ in (3.6b) ${\ds s^{\beta_2}}$ is negligibly small
in comparison to ${\ds s^{\beta_1}}$for $s \to 0^+$ and, viceversa,
   ${\ds s^{\beta_1}}$ is negligibly small
in comparison  ${\ds s^{\beta_2}}$for $s \to +\infty$.

\subsection{Fractional diffusion of uniformly distributed order}
Second, we  consider the choice
$$
p(\beta ) =1\,, \q 0 <\beta < 1\,.  \eqno(3.9)$$
For  \underbar{the R-L case} we have
$$A(s) = s \! \int_0^1 \! s^{-\beta} \, d\beta  = \frac{s-1}{\log s}\,, \eqno(3.10a)$$
hence, in virtue of (3.3a),
$$
\widetilde {\sigma^2}(s) = {2}\,\lt[ \frac{1}{s\log s}- \frac{1}{s^2\log s}\rt]\,. \eqno(3.11a)$$  
Then, by inversion, see Appendix C, Eqs (C.15)-(C.16), we get
$$
\sigma ^2(t) =
2 \left  [ \nu (t,0)- \nu (t,1)\right]  
  \sim
 \cases{
 2/\log(1/t)\,, &  $t \to 0\,,$ \cr
2t/\log t\,,  &  $t \to \infty\,,$}
\eqno(3.12a)$$
where 
$$\nu(t, a) := \int_0^\infty \, \frac {t^{a+\tau}} {\Gamma(a+ \tau +1)}\, d\tau\,,\q a > -1\,,$$
denotes  a special function  just introduced in Appendix C along with its Laplace transform. 
\vsp
For \underbar{the C case}    we have
$$B(s) = \int_0^1 \! s^\beta \, d\beta  = \frac{s-1}{\log s}\,, \eqno(3.10b)$$
hence, in virtue of (3.3b),
\vsp
\begin{figure}[h!]
\begin{center}
\includegraphics[scale=0.40]{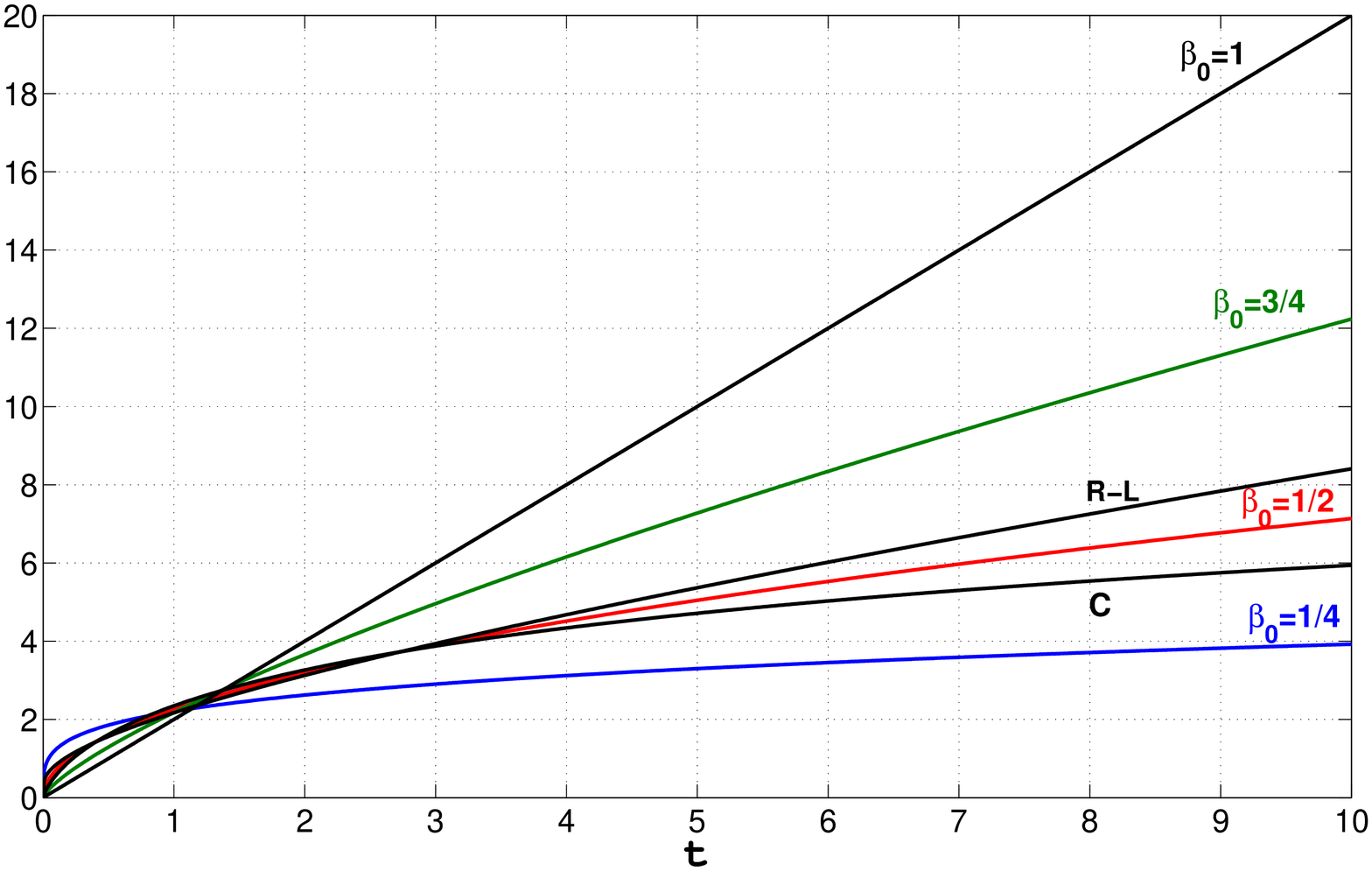}
\includegraphics[scale=0.40]{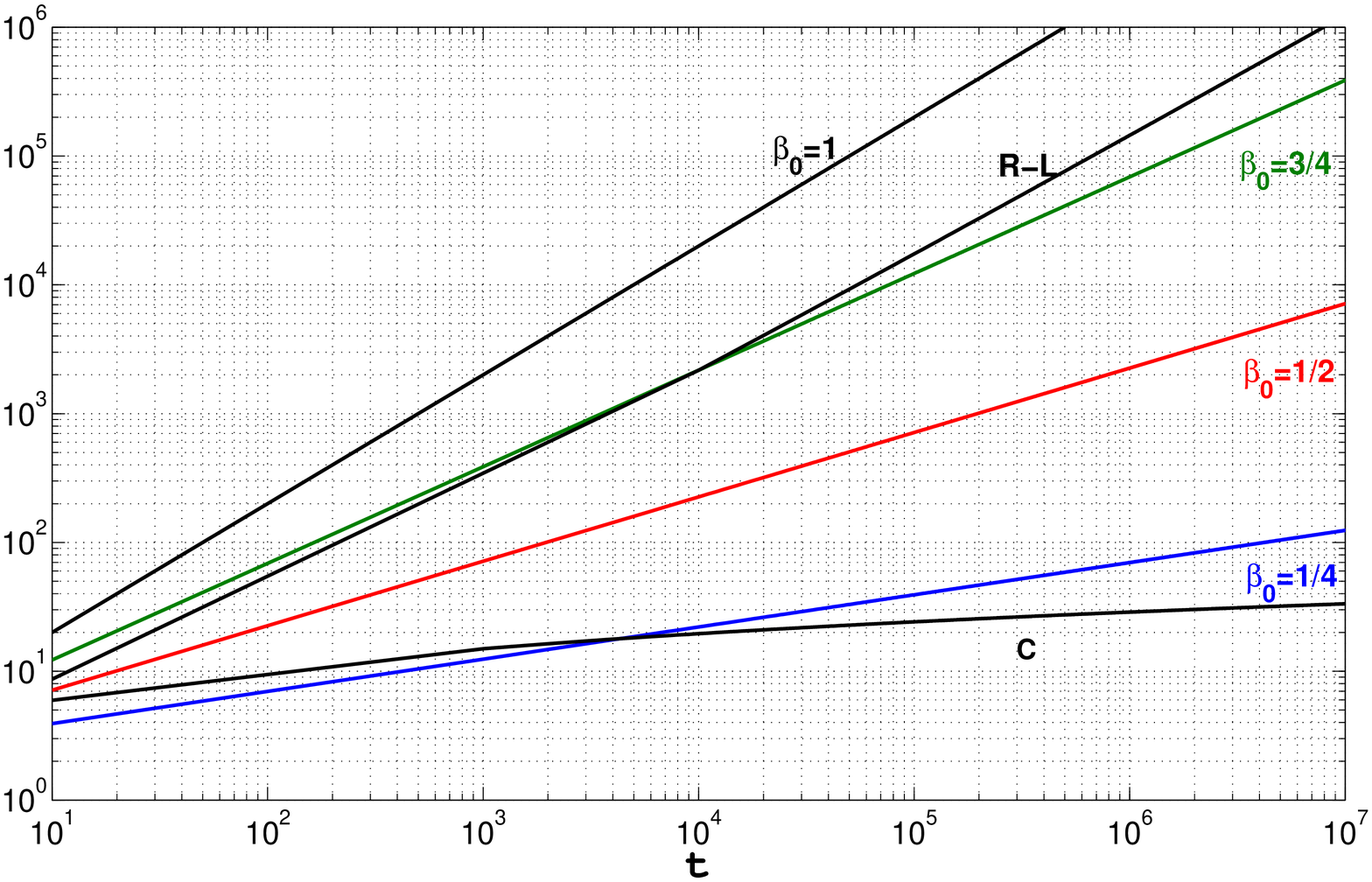}
\end{center}
\vskip -0.5truecm
 \caption{ Variance versus $t$ 
 for the uniform order distribution 
 in  R-L and C forms compared
  with some single order case.
  Top: $0 \le t \le 10$ (linear scales);
  Bottom: $10^1\le t\le 10^7$ (logarithmic scales).
  }
 \end{figure}
 $$
\widetilde {\sigma_*^2}(s) = \frac{2}{s}\, \frac{\log s}{s-1} \,. \eqno(3.11b)$$ 
Then, by inversion, see Appendix C, Eqs. (C.11), (C.14), and compare with 
Eqs. (23)-(27) in \cite{ChechkinGorenfloSokolov PRE02}
$$
\sigma _*^2(t) =
2 \left  [ \log t + \gamma + \e^{\, \ds t}\, \EE_1(t)\right  ] 
 \sim \,
 \cases{
 2t\, \log (1/t), &  $t \to 0,$ \cr
2\, \log (t),  &  $t \to \infty\,,$}
\eqno(3.12b)$$
where 
$$\EE_1 (t): = \int_{t}^\infty \!\! \frac{\;\e^{-u}}{ {\ds u}}\, du
  =  \e^{\ds -t} \,
 \int_{0}^\infty \!\! \frac{\;\e^{-u}}{ {u+t}}\, du  $$ 
 denotes
the exponential integral  function recalled in Appendix C,
 and $\gamma = 0.57721...$ is the so-called Euler-Mascheroni constant.
 \vsp
For the uniform distribution we find it instructive to compare the time evolution
of the variance
 for the R-L and C forms with that corresponding to  a few of  single orders.  
In Fig. 1-Top  we consider  moderate times ($0\le t\le 10$) using linear scales,
whereas in Fig. 1-Bottom large times ($10^1\le t\le 10^7$) using logarithmic scales.

\section{Evaluation of the fundamental solutions}

\subsection{The two strategies}
In order
to determine the fundamental  solutions $u(x,t)$
and $u_*(x,t)$
in the space-time domain we can follow two
alternative  strategies related to the
order in carrying out the Fourier-Laplace  in (2.6a)
and (2.6b) 
\\
(S1) : invert  the Fourier transforms
getting $\widetilde{u} (x,s)$, $\widetilde{u}_* (x,s)$
   and   then invert the remaining  Laplace transforms;
\\
(S2) : invert  the Laplace transform getting $\widehat{u} (\kappa ,t)$,
$\widehat{u}_* (\kappa ,t)$
and then invert the remaining  Fourier transform.
\vsp
Before considering the general case of time-fractional diffusion
of distributed order, we prefer to briefly recall  
the determination of the fundamental solution $u(x,t)$ (common for
both the R-L and C forms) for the single order case.

\subsection{The single order diffusion}

For the time-fractional diffusion equation of single order $\beta _0$
the strategy (S1) yields the Laplace transform 
$$ \widetilde{u} (x,s) =
 {s^{\beta _0 /2- 1} \over 2 }\,\e^{\ds - |x| s^{\beta _0 /2}}\,,
\q 0<\beta _0\le 1\,.
 \eqno(4.1) $$
Such strategy was adopted 
by Mainardi
\cite{Mainardi WASCOM93,Mainardi CSF96,Mainardi CISM97}
to obtain the  Green function  in the form
$$ u (x,t) = t^{-\beta _0/2}\,
       U \left (|x|/t^{\beta _0/2}\right )\,,
\q -\infty < x <+\infty\,,  \q t\ge 0\,,
\eqno(4.2)
$$
where the variable
 $ X:= x/t^{\beta _0/2}$ acts as {\it similarity variable}
and the function $U(x) := u(x,1)$ denotes
the {\it reduced Green function}.
Restricting from now on  our attention to $x\ge 0$,
the solution turns out   as
$$
\begin{array}{ll}
U(x) &={\ds \frac{1}{2}}\, M_{\beta _0/2} (x)
  = 
   {\ds \frac{1}{2}}\,    {\ds \sum_{k=0}^{\infty}\,
  \frac{(-x)^k}{ k!\,\Gamma[-\beta _0 k/2 + (1-\beta _0/2)]}}  \\
& = {\ds \rec{2\pi}\, \sum_{k=0}^{\infty}\,\frac{(-x)^k }{ k!}\,
  \Gamma[(\beta _0 (k+1)/2]  \,\sin [(\pi \beta _0(k+1)/2]}
\,,
\end{array}
\eqno(4.3)$$
where $M_{\beta _0/2} (x)$
is an an entire
 transcendental function (of order $1/(1-\beta _0/2)$)
of the Wright type,
see also \cite{GoLuMa 99,GoLuMa 00,Mainardi-Pagnini AMC03} and  \cite{Podlubny 99}.
\vsp
\begin{figure}[h!]
\begin{center}
\includegraphics[scale=0.40]{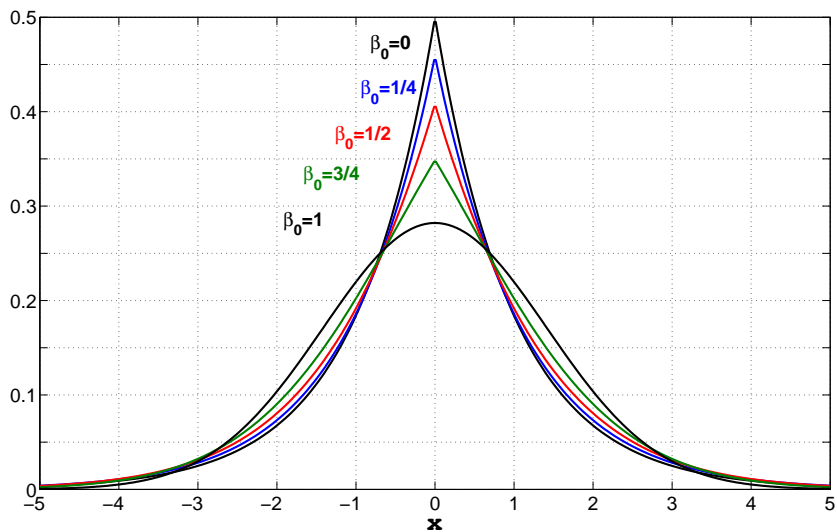}
\end{center}
\vskip -0.5truecm
 \caption{The reduced Green function 
 $U(x)= \frac{1}{2} M_{\beta_0/2}(x))$
 versus $x$ (in the interval $|x|\le 5$),
 for $\beta_0=0, 1/4, 1/2, 3/4, 1$.}
 \end{figure}
\vsp
Since the  fundamental solution has the peculiar property to be  self-similar
it  is sufficient to consider  the reduced Green function $U(x)$.
In Fig. 2 we  show the graphical
representations of $U(x)$ for different orders 
ranging from $\beta_0=0$, for which we recover the Laplace density
$$  U(x) = \frac{1}{2}\, \e^{\,\ds -|x|}\,, \eqno(4.4)$$
 to $\beta_0=1$,
for which we recover the Gaussian density (of variance $\sigma^2=2$)
$$ U(x)= \frac{1}{2\sqrt{\pi }}\, \e^{-\ds x^2/4}\,. \eqno(4.5)
$$
The  Strategy (S2): yields the  Fourier transform.
$$ \widehat{u}(\kappa ,t) =  E_{\beta _0} \left(-\kappa ^2 t^{\beta _0} \right)\,,
\q 0<\beta _0\le 1\,,    \eqno(4.6) $$
where $E_{\beta _0}$ denotes the  Mittag-Leffler function,
see Appendix B.
The strategy (S2) has been followed
 by Gorenflo et al. \cite{GoIsLu FCAA00} and
by Mainardi et al. \cite{MaLuPa FCAA01,MaPaSa JCAM05}
to obtain the  Green functions of the more general
space-time-fractional diffusion equations (of single order),
and requires to invert the Fourier transform
by using the machinery of the Mellin  convolution
and the   Mellin-Barnes integrals. 
Restricting ourselves here to recall the final results,
 the reduced Green function
for the time-fractional diffusion equation
 now appears, for $x \ge 0$,  in the form:
$$ 
U(x) = {\ds \rec{\pi}\int_{0}^{\infty}
\!\! \cos\,(\kappa  x)  \,
  E_{\beta _0} \left (-\kappa^2\right )\, d\kappa}
 = {\ds \frac{1}{2x}\, \frac{1}{2\pi i}\,
   \int_{\sigma-i\infty}^{\sigma +i\infty}
  \!\!\frac{\Gamma(1-s)}{\Gamma (1- \beta_0 s/2)}
 \, x^{\,\ds s}\,  ds}\,,
 \eqno (4.7) $$
with $0 <\sigma< 1$.
By    solving the  Mellin-Barnes integrals using the residue theorem,
we arrive at the same power series (4.3) of the $M$-Wright function.
Both strategies allow us to prove that the Green function is
non-negative and normalized, so it can be interpreted
as a spatial probability density evolving in time with the similarity law
(4.2).
\vsp
For readers' convenience we  like to mention other papers dealing with the fundamental solutions
of fractional  diffusion equations (of single ordrer); a non exhaustive list includes
\cite{AnhLeonenko JSP01,Eidelman-Kochubei JDE04,Hanyga TFD-PRS02,Hilfer BOOK00,Kochubei DE90,
Metzler PhysA94,Metzler-Klafter PhysRep00,
Saichev-Zaslavsky CHAOS97,SchneiderWyss 89,Zaslavsky PhysRep02} and references therein. 
\subsection{The distributed order diffusion}
Similarly with the single order diffusion, also for the cases of distributed order
we can follow either strategy (S1) or strategy (S2). 
In contrast with  previous papers of our group where we have  followed
the strategy (S2), 
see \cite{Mainardi-et-al SPRINGER-MME06,Mainardi-Pagnini JCAM06,Mainardi-Pagnini-Gorenflo AMC06},
here we follow   the  strategy (S1).
This choice implies to recall 
the Fourier transform pair (a straightforward exercise
in complex analysis based on residue theorem and  Jordan's lemma)
$$  {c  \over  d + \kappa ^2}
  \,\stackrel{\F} {\leftrightarrow}\,
\frac{c}{2 d^{1/2}}\,\e^{\ds - |x| d^{1/2}}\,,
 \q d >0 \,.
 \eqno (4.8)$$
 In fact we recognize by comparing (4.8) with (2.6a)-(2.6b) that for the RL and C forms we have
 $$
 \hbox{R-L}:\; \cases{c= c(s):= 1/A(s)\cr
            d= d(s):= s/A(s)} \quad
\hbox{C}:\; \cases{c = c(s) := B(s)/s\,, \cr
         d=d(s):=B(s) }
		 \eqno(4.9) $$	 		
Now we have to invert the Laplace transforms obtained inserting (4.9)
in the R.H.S of (4.8).
\vsp
For \underbar{the R-L case} we have:
$$ \widetilde u(x,s) =  \frac{1}{2[sA(s)]^{1/2}} \, \exp \left\{-|x| [s/A(s)]^{1/2}\right\}\,. \eqno(4.10a)$$
For \underbar{the C case} we have:  
$$ \widetilde u_*(x,s) =  \frac{[B(s)]^{1/2}}{2 s} \, \exp \left\{-|x| [B(s)]^{1/2}\right\}\,. \eqno(4.10b)$$
Following a standard procedure in complex analysis, the Laplace inversion requires
the integration along the borders of the negative real semi-axis in the $s$-complex cut plain;
in fact this semi-axis, defined by $s= r \e^{\pm i\pi}$ with $r>0$ turns out  the branch-cut  
common for the functions $s^{1-\beta}$ (present in $A(s)$for the RL form) and $s^\beta$ 
(present in $B(s)$ for the C form).
Then,  in virtue of  the Titchmarsh theorem on Laplace inversion, we get
the representations in terms of real integrals of Laplace type.
\vsp
For \underbar{the R-L case}  we get
$$   {u} (x,t) = -\rec{\pi }\,
\int_0^\infty \! \e^{-rt} \,  \hbox{Im}
\left\{\widetilde {u} \left(x,r \e^{i\pi }\right)\right\}
\, dr\,, \eqno(4.11a) $$
where, in virtue of (4.10a),  we must know $A(s)$
along the ray $s=r\,\e^{i\pi }$ with $r>0$.
We write
$$ A \lt(r\, \e^{\,\ds i \pi}\rt) =
\rho \,\cos (\pi \gamma)
+ i \rho \sin (\pi \gamma)\,, \eqno(4.12a)$$
where
$$
\cases{
{\ds  \rho =\rho (r) =\left\vert A\lt(r \,\e^{i\pi }\rt) \right\vert}\,, \cr
{\ds \gamma = \gamma (r) =
\rec{\pi}\,\hbox{arg}  \left[A\lt (r \,\e^{i\pi }\rt)\right]}\,.
} \eqno(4.13a)$$
 Similarly for the \underbar{C case}
 we obtain
$$   
{u}_* (x,t) \!= \! -\rec{\pi }
\int_0^\infty \!\! \e^{-rt} \,  \hbox{Im}
\left\{\widetilde {u}_* \left(x,r \e^{i\pi }\right)\right\}
\, dr\,, \eqno(4.11b) $$
where, in virtue of (4.10b),  we must know $B(s)$
along the ray $s=r\,\e^{i\pi }$ with $r>0$.
We write
$$ B \left(r\, \e^{\,\ds i \pi}\right) =
\rho _* \,\cos (\pi \gamma _*)
+ i \rho _* \sin (\pi \gamma _*)\,, \eqno(4.12b)$$
where
$$
\cases{
{\ds  \rho_* =\rho_* (r) =\left\vert B\lt(r \,\e^{i\pi }\rt) \right\vert}\,, \cr
{\ds \gamma_* = \gamma_* (r) =
\rec{\pi}\,\hbox{arg}  \left[B\lt (r \,\e^{i\pi }\rt)\right]}\,.
} \eqno(4.13b)$$
As  a consequence we formally write the  required  fundamental solutions as
$$ u(x,t) = \int_0^\infty \e^{-rt}\, P(x,r)\,dr \,,
\quad  
P(x,r) =  -\rec{\pi }\,
  \hbox{Im}
\left\{\widetilde {u} \left(x,r \e^{i\pi }\right)\right\}
\,, \eqno(4.14a) $$
and
$$ u_*(x,t) = \int_0^\infty \e^{-rt}\, P_*(x,r)\,dr \,, 
\quad  
P_*(x,r) =  -\rec{\pi }\,
  \hbox{Im}
\left\{\widetilde {u}_* \left(x,r \e^{i\pi }\right)\right\}\,,
\eqno(4.14b)$$
where the functions  $P(x,r)$ and $P_*(x,r)$ must be derived by using
Eqs. (4.10a)-(4.14a) and Eqs. (4.10b)-(4.14b), respectively.
We recognize that, in view of the transformation (2.7), the expressions of $P$ and $P_*$
are related to each other by the  transformation
$$    \rho _* (r) \Longleftrightarrow r/\rho(r)\,,\quad \gamma _*(r)  \Longleftrightarrow  1-\gamma(r)\,.
\eqno(4.15)$$
 We then  limit ourselves to provide the explicit expression \underbar{for the C form}
 $$
 \begin{array}{ll}
  P_*(x,r) 
&= {\ds \frac{1}{2\pi r}}\,
\hbox{Im}\left\{
\rho _*^{1/2}  \, \e^{\, \ds i\pi\gamma _*/2} 
 \,\exp \left[ \ds \,-\, \e^{\, \ds i\pi\gamma _*/2}\rho _*^{1/2}x  \right]\right\} \\ \\ 
&= {\ds \frac{1}{2\pi r}}\,
\rho _*^{1/2} \,
{\ds \e^{\ds \, -\rho_*^{1/2}x \cos(\pi\gamma _*/2)}}
\, \sin \left[\pi \gamma _*/2 - \rho_*^{1/2} x \sin(\pi\gamma _*/2)\right]\,.
\end{array}
\eqno(4.16)
$$ 
For the \underbar{R-L form} the corresponding expression of $P(x;r)$ is obtained
 from (4.16) by applying the transformation (4.15). 
\vsp
We note that the fundamental solutions found in this subsection 
are equivalent to those  obtained by the Authors in \cite{Mainardi-et-al SPRINGER-MME06}
by following the strategy (S2) after  a lengthy manipulation of Mellin-Barnes integrals.
\vsp
Herafter we exhibit some plots of the fundamental solutions for the two case studies
considered in subsection 3.2 in order to point out the remarkable difference
between the R-L and the C forms.
\newpage
 \subsection{Plots of the  fundamental solutions}
 For the case of two orders, we chose 
 $\{\beta_1=1/4, \beta_2=1\}$ in order to  contrast the 
 evolution
 of the fundamental solution  for the R-L and the C forms.
\begin{figure}[h!]
\begin{center}
\includegraphics[scale=0.40]{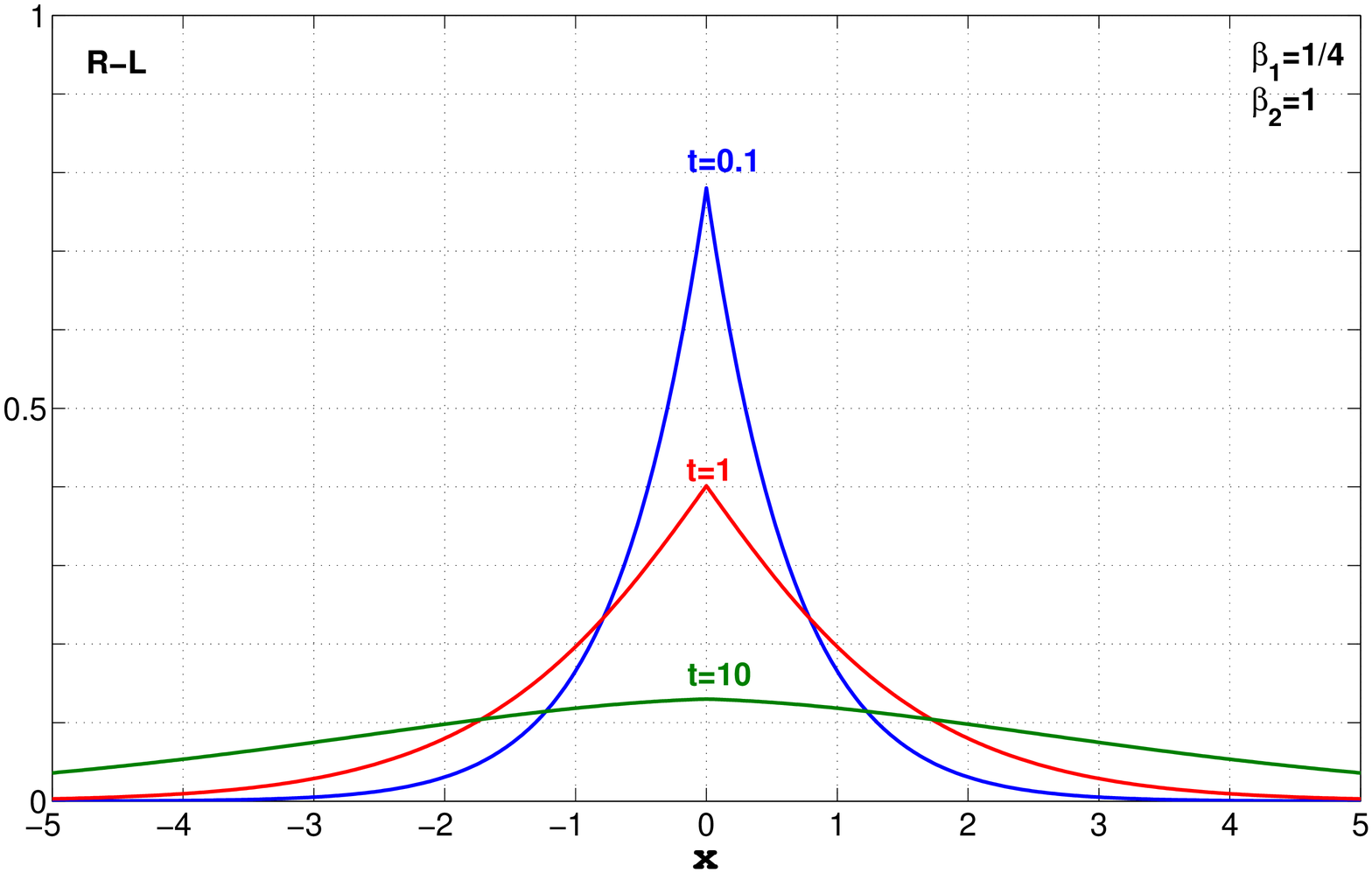}
\vskip -0.2truecm
\includegraphics[scale=0.40]{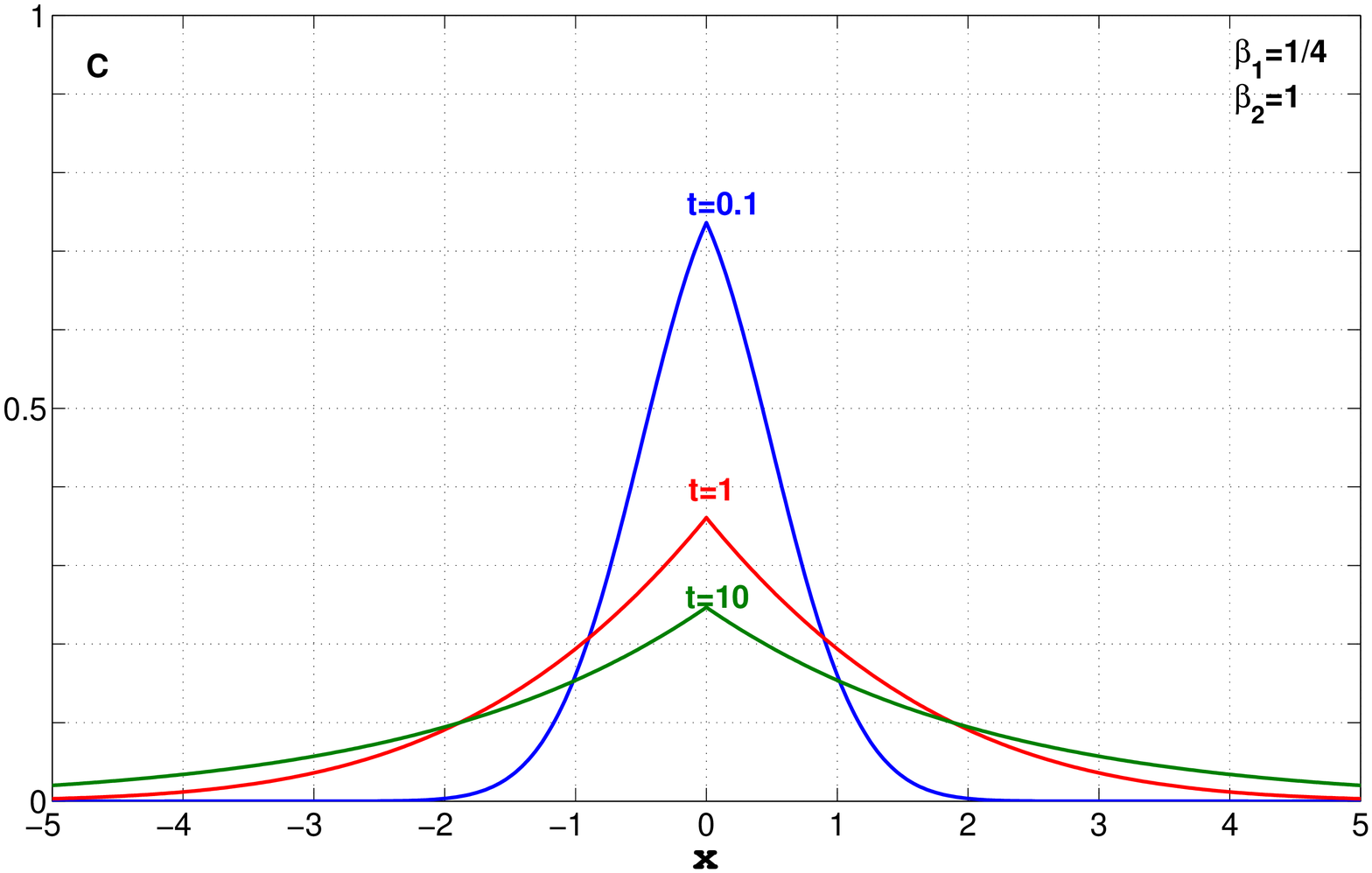}
\end{center}
\vskip -1.00truecm
 \caption{The fundamental solution
 versus $x$ (in the interval $|x|\le 5$),
 for the double-order  distribution $\{\beta_1=1/4, \beta_2=1\}$ at times  $t= 0.1,1, 10$.
 Top: R-L form; Bottom: C form.}
 \end{figure}
\begin{figure}[h!]  
\begin{center}
\includegraphics[scale=0.40]{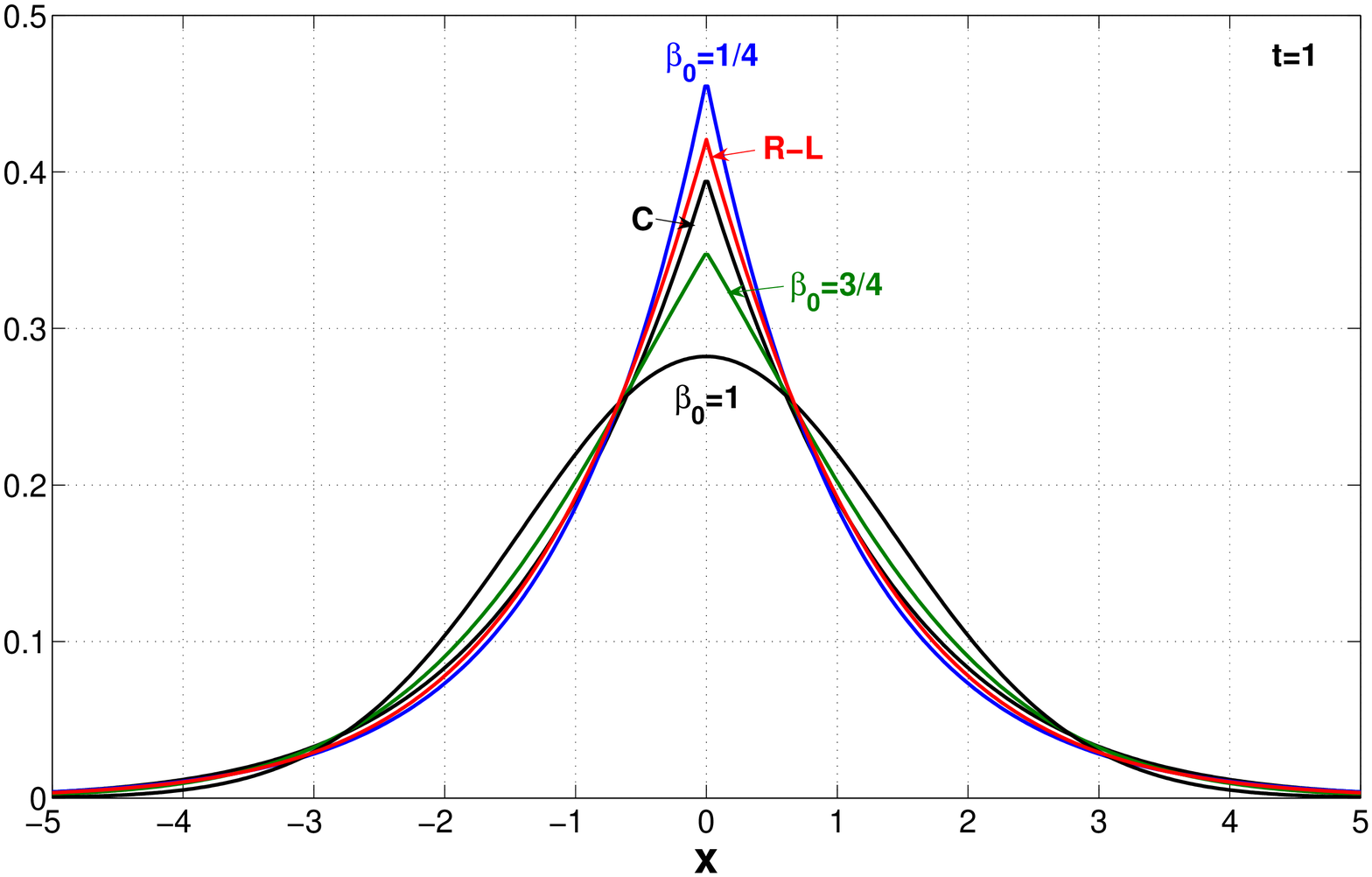}
\vskip -0.2truecm
\includegraphics[scale=0.40]{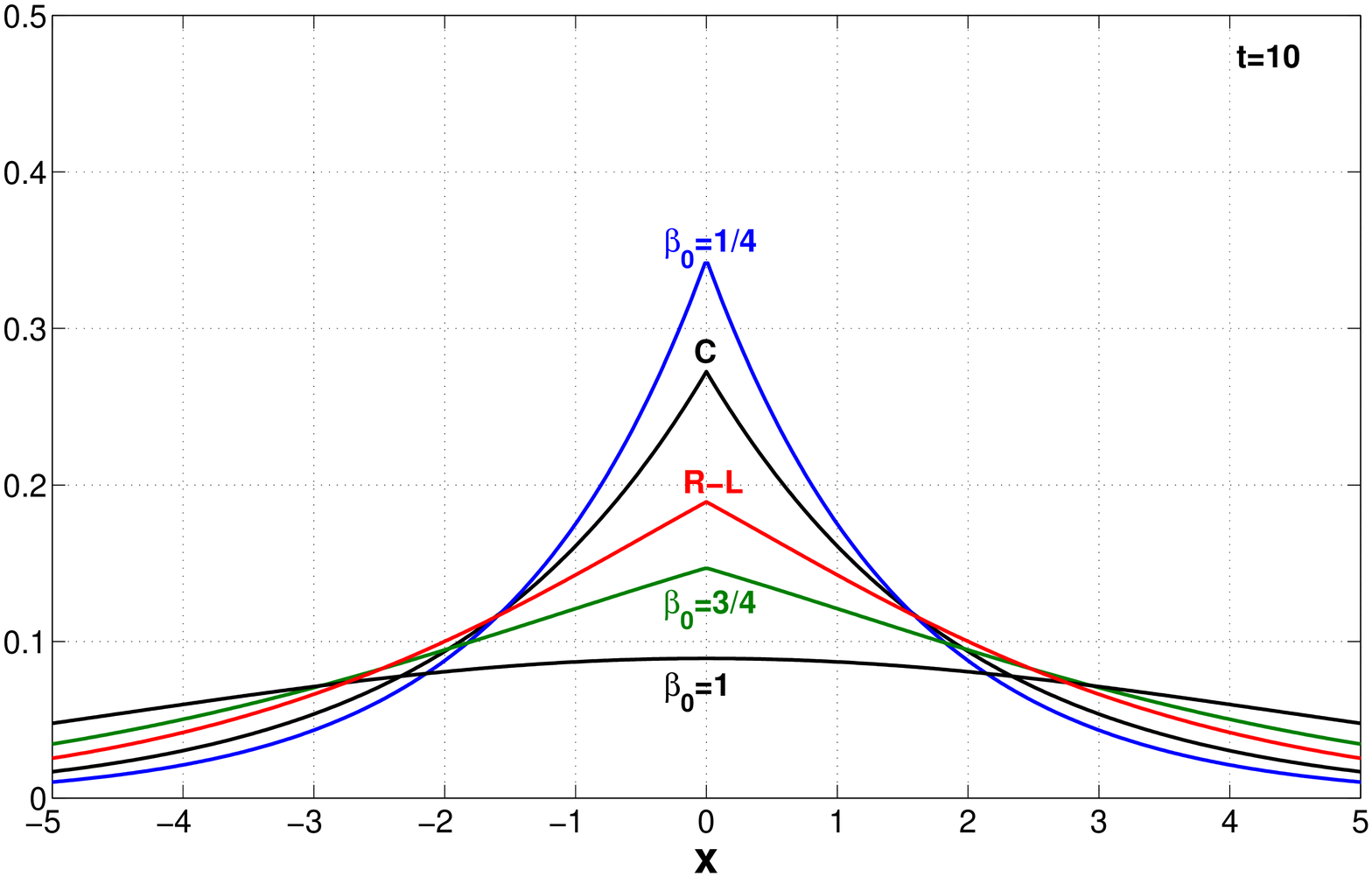}
\end{center}
\vskip -1.00truecm
 \caption{The fundamental solutions
 versus $x$ (in the interval $|x|\le 5$),
 for the uniform order distribution in R-L and C forms
 compared with the solutions for some cases of single order. 
 Top: $t=1$; 
 Bottom: $ t=10$.}
 \end{figure}
\vsp
 In Fig. 3    we exhibit the plots of the corresponding  solution
versus $x$ (in the interval $|x|\le 5$),
 at different times, selected as $t=0.1$,
 $t=1$ and $t=10$.
 In this limited spatial range we can note how the time evolution
 of the $pdf $depends  on the different time-asymptotic behaviour of the variance,
  for the two forms, as  stated in Eqs. (3.12a)-(3.12b), respectively.
\vsp
For the uniform distribution, we find it instructive to compare in Fig. 4
the solutions corresponding to R-L and C forms
with the solutions of the   fractional diffusion of a single order
$\beta_0 = 1/4,  3/4, 1$ at fixed times, selected as
 $t= 1,10$. We have skipped  $\beta _0=1/2$ 
 for a better view of the plots.
\section{ Conclusions}
We have investigated the time fractional diffusion equation with (discretely or
continuously) distributed order between 0 and 1  
in the Riemann-Liouville and in the Caputo forms, 
providing the  Fourier-Laplace representation
of the corresponding fundamental solutions.
Except for the case of a single order, for which
the two forms are equivalent with
a self-similar fundamental solution,  
for a general order distribution the equivalence
and the self-similarity are lost.
In particular  the asymptotic behaviour of the fundamental
solution and its variance  at small and large times strongly depends on the
selected approach.   
We have considered
  two simple  but noteworthy case-studies 
 of distributed order, namely 
 the case of a superposition of two different orders $\beta_1$ and $\beta_2$
and the case of a uniform order distribution.
In the first case one of the orders dominates the time-asymptotics
near zero, the other near infinity,
but $\beta_1$ and $\beta_2$
change their roles when switching from the R-L form to the C form
of the time-fractional diffusion. 
The asymptotics for uniform order density is remarkably different,
the extreme orders now being (roughly speaking) 0 and 1.
We now meet super-slow and slightly super-fast time behaviours
of the variance near zero and near infinity, again with the interchange of behaviours
between the R-L and C form.  
We have clearly pointed out the above effects
with the figures  in sub-section 3.3,
in particular the extremely slow growth of the variance as
$t \to \infty$ for the C form.
After the analysis of the variance, that in practice requires only the inversion  
of a Laplace transform, we have considered the task of the double inversion of
the Laplace-Fourier representation. 
For a general order distribution
we were able to  express the fundamental
solution  in terms of a Laplace integral in time
with a kernel which depends on space and order distribution in a
simple form, see Eqs. (4.14)-(4.16).
For the two case studies
the plots  of the fundamental solutions (reported in sub-section  4.4)
have shown 
their dependence on the different asymptotic behavior of the corresponding variance.

\newpage 

 \section*{Acknowledgements}
  This work, carried out in the framework of a  research
  project for {\it Fractional Calculus Modelling} 
 [URL: {\tt www.fracalmo.org}],
 is  based on a plenary lecture given by F.M. at the
 second  IFAC Workshop on {Fractional Differentiation 
  and its  Applications} (FDA06), Porto (Portugal) 19-21 July 2006.
F.M. is grateful to  the organizing committee of FDA06 
for this invitation. 

\section*{Appendix A: The two fractional derivatives}

The purpose of this Appendix is to clarify for the interested reader
the main differences between
the {\it Riemann-Liouville} (R-L) fractional derivative
and the {\it Caputo}  (C) fractional derivative\footnote{
This form is referred to Caputo who used it 
in the late sixties of the past century, see \cite{Caputo 67,Caputo 69}.
Soon later this derivative was adopted by Caputo and Mainardi
 in the framework of the theory of {\it Linear Viscoelasticity},
 see \cite{CaputoMaina 71}.
 It was mainly  with  the 1997 CISM chapter by Gorenflo and Mainardi
 \cite{GorMai CISM97} and with the 1999 book by Podlubny \cite{Podlubny 99}
    that such form was popularized.}
for well-behaved functions $f(t)$ with $t\ge 0$, exhibiting 
a finite limit $f(0^+)$ as $t \to 0^+\,. $
Denoting with  $\mu \in (0,1]$ the order,    
the R-L derivative is defined as 
$$
 _tD^\mu  \,f(t) := \cases{
  {\ds	\rec{\Gamma(1-\mu)}}\,
  {\ds {d \over dt}}\,
 {\ds \int_0^t
    {f(\tau)  \over (t-\tau )^\mu}\,d\tau} \,,
  & $\; 0<\mu  <1\,,$\cr\cr
 {\ds {d \, f(t) \over dt}}\,, & $\; \mu =1\,.$\cr}
\eqno(A.1) $$ 
and the C derivative as 
$$
 _tD_*^\mu  \,f(t) := \cases{
  {\ds	\rec{\Gamma(1-\mu)}}\,
 {\ds \int_0^t
    {f^\prime(\tau)  \over (t-\tau )^\mu}\,d\tau} \,,
  & $\; 0<\mu  <1\,,$\cr\cr
 {\ds {d \, f(t) \over dt}}\,, & $\; \mu =1\,.$\cr}
\eqno(A.2) $$
The two  fractional derivatives  are related to
the Riemann-Liouville fractional  integral as follows.
The Riemann-Liouville fractional  integral    is
$$ _tJ^\mu \, f(t) :=
\rec{\Gamma(\mu )}\,\int_0^t
    {f(\tau)  \, (t-\tau )^{\mu -1}}\,d\tau \,,\q \mu >0\,,
  \eqno(A.3) $$
(with the convention $\,_tJ^0 \, f(t) = f(t)$)
and is known to satisfy the semigroup property
$ _tJ^\mu\, _tJ^\nu = \,_tJ^{\mu+ \nu }\,, $
with $\mu, \nu >0\,.\,$
For any $\mu >0$
the Riemann-Liouville fractional derivative is
defined as the {\it left} inverse of the corresponding
fractional integral
(like  the derivative of any integer order), namely
$   \, _tD^\mu \, _tJ^{\mu}\, f(t)= f(t)\,.$
Then for  $\mu \in (0,1] $ we have
 $$   _tD^\mu\, f(t) := \,_tD^1\, _tJ^{1-\mu}\, f(t)\,,\q
      _tD_*^\mu f(t) := \,_tJ^{1-\mu} \, _tD^1\, f(t)\,,
\eqno(A.4)$$
$$
_tJ^\mu \, _tD_*^{\mu}\, f(t)=
 \,_tJ^\mu\, _tJ^{1-\mu}\, _tD^1\, f(t) =
  \,_tJ^1\, _tD^1\, f(t) = f(t) - f(0^+)\,.
\eqno(A.5)
$$
Recalling the rule
$$ _t D^{\mu }\, t^{\gamma}=
   {\Gamma(\gamma +1)\over\Gamma(\gamma +1-\mu)}\,
     t^{\gamma-\mu }\,,
 \q \mu  \ge 0\,,
  \q \gamma >-1\,, \eqno(A.6)
$$
it turns out  for  $0 <\mu < 1\,,$
$$ _tD_*^\mu	\,f(t)	\, = \, _tD^\mu  \,\left[ f(t) -
   f(0^+) \right] =
\, _tD^\mu \, f(t) -
      f(0^+) \,
{t^{-\mu }\over \Gamma(1-\mu)}\,.
 \eqno(A.7) $$
Note that for $\mu=1$ the two types of fractional derivative coincide,
the constant $f(0^+)$ playing no role.
\vsp
The {\it Caputo} fractional derivative
represents a sort of regularization in the time origin for the
{\it Riemann-Liouville} fractional derivative
and  satisfies the  relevant property
of being zero when applied to a constant.
\vsp
Let us now consider the behaviour of  the above derivatives 
of non integer order with respect to
the Laplace transformation\footnote{%
The Laplace transform of a well-behaved function $f(t)$
 is defined as
 $$\widetilde f(s) =
\L \left\{ f(t);s\right\}
 := {\ds \int_0^{\infty}} \! \e^{\ds \, -st}\, f(t)\, dt\,, \;s \in \CC\,.$$
 We recall that under suitable conditions the Laplace transform of the 
 first derivative of $f(t)$ is given by
 $$ \L \left\{ _tD^1 \,f(t) ;s\right\} =
      s \,  \widetilde f(s) - f(0^+)\,,
\quad f(0^+) :=  \lim_{t\to 0^+}\, f(t)\,.$$   
 }.
\vsp
For the Riemann-Liouville
derivative of non-integer order $\mu$  we have
$$
  \L \left\{ _tD^\mu  \, f(t);s\right\} \!=\!
     s^\mu \,  \widetilde f(s) - g(0^+),
\;  g(0^+) \!:=\! {\ds \lim_{t\to 0^+} \,_tJ^{1-\mu} f(t)},  \; 0<\mu<1.
\eqno(A.8)$$
For the Caputo derivative of non-integer order $\mu$  
we have
$$ \L \left\{ _tD_*^\mu \,f(t) ;s\right\} \!=\!
      s^\mu \,  \widetilde f(s)
   - s^{\mu-1}\, f (0^+) \,,
\; f(0^+) \!:=\! {\ds \lim_{t\to 0^+}\, f(t)\,,} \; 0<\mu<1\,.   
\eqno(A.9)$$
 Thus the rule  (A.8) is     more cumbersome to be used
than (A.9) since it requires initial values  concerning
an extra function $g(t)$ related to the given $f(t)$  
through a  fractional integral. However,
when the limiting   value $f(0^+)$
is finite we can see that $g(0^+)$
is vanishing so  
the formula (A.8)  simplifies into
$$ \L \left\{ _tD^\mu  \, f(t);s\right\} =
      s^\mu \,  \widetilde f(s) \,,\quad 0<\mu<1\,. \eqno(A.8')$$
\vsp	  
For further reading
on the theory and applications of fractional calculus
we recommend
to consult in addition to the
well-known books by Samko, Kilbas \& Marichev \cite{SKM 93},
by Miller \& Ross \cite{Miller-Ross 93},
by Podlubny \cite{Podlubny 99},
 those appeared in the last few years,
by  Kilbas, Srivastava \& Trujillo \cite{Kilbas-et-al BOOK06},
by Magin \cite{Magin BOOK06},
by West, Bologna \& Grigolini \cite{West BOOK03},
and by Zaslavsky \cite{Zaslavsky BOOK05}.


\section*{Appendix B: The Mittag-Leffler functions}

\subsection*{B.1 \ The classical Mittag-Leffler function}
Let us recall that the Mittag-Leffler function $E_{\mu}(z)$ 
(with $\mu>0$)
is an entire  transcendental function of order $1/\mu $, defined
in the complex plane by the power series
$$ E_{\mu} (z) :=
    \sum_{k=0}^{\infty}\,
   {z^{k}\over\Gamma(\mu\,k+1)}\,, 
 \q z \in\CC\,. \eqno(B.1)  $$
 It was introduced and studied by the Swedish mathematician
 Mittag-Leffler at the beginning of the XX century
 to provide a noteworthy example of entire function 
 that generalizes the exponential  (to which it reduces
 for $\mu=1$).
For  details on this function  we refer \eg to
\cite{Djrbashian 66,Erdelyi HTF,GorMai CISM97,Kilbas-et-al BOOK06,MaiGor JCAM00,Podlubny 99,SKM 93}.
In particular we  note that the function $E_{\mu} (-x)$ ($x\ge 0$)
 turns a completely monotonic function of $x$ if $0< \mu \le 1$.
  This property is still valid
 if we consider the variable   $x = \lambda \,t^\mu$ where
 $\lambda$ is a positive constant. Thus  the function
$E_\mu (- \lambda t^\mu ) $ preserves the   {\it complete  monotonicity}
of the exponential $\exp(- \lambda t)$: indeed,
for $0<\mu<1$  it
 is represented in terms of a real Laplace transform (of a real parameter $r$)
of a non-negative   function (that we refer  to as the spectral function)
$$ 
E_\mu (- \lambda t^\mu)  = 
   \frac{1} {\pi}
   \int _0^\infty \!  {\e^{\,\ds -r t}}\,
   \frac{ \lambda r ^{\mu-1}    \,\sin (\mu \pi) }
    {\lambda^2 + 2 \lambda \, r ^\mu \,\cos(\mu\pi)+ r^{2\mu} }
        \,dr\,. 
  \eqno(B.2)$$
  We note that as $\mu \to 1$ the spectral function tends
  to the generalized Dirac function $\delta(r-\lambda)$.
 \vsp
  We  note that the Mittag-Leffler function (B.2)
  starts at $t =0$ as a stretched exponential and  
   decreases for  $t \to \infty$  
like a power  with exponent $-\mu $:
$$  E_\mu (-\lambda t^\mu ) \sim 
\cases{
{\ds 1 - \lambda \,\frac{t^\mu}{ \Gamma(1+\mu)} } 
 \sim  {\ds \exp \left\{- \frac{\lambda t^\mu}{ \Gamma(1+\mu)} \right\} }\,,
 & $t \to 0^+\,,$ \cr\cr
{\ds \frac{t^{-\mu }}{ \lambda \,\Gamma(1-\mu)} }\,,
 & $t \to \infty\,.$ \cr
 }
\eqno(B.3)$$
The  noteworthy results (B.2) and (B.3) can also be derived from 
the Laplace transform pair
$$ \L\{ E_\mu(-\lambda t^\mu);s\} = \frac{s^{\mu -1} }{s^\mu + \lambda}\,.\eqno(B.4)$$
In fact  it it sufficient to apply the Titchmarsh theorem
($s = r \e^{i\pi}$) for deriving (B.2) and
the Tauberian theory  ($s\to \infty$ and $s \to 0$) for deriving (B.3).
\vsp
If $\mu =1/2$
we have for $t\ge 0$:
$$ 
E_{1/2} (-\lambda \sqrt{t})  =
    \e^{\ds \,\lambda^2 t}\, \hbox{erfc} (\lambda \sqrt{t}) 
\sim 1/(\lambda \sqrt{\pi \,t})\,,\; t\to \infty \,,
 \eqno(B.5)$$
where $ \, \hbox{erfc}\,$ denotes the {\it complementary error}
function, see e.g. \cite{AS 65}.   
\subsection*{B.2 \ The generalized Mittag-Leffler function}
 The Mittag-Leffler function  in two parameters
$E_{\mu,\nu}(z) $ ($ \Re \{\mu\} >0$,  $\, \nu \in \CC$)
is  defined 
by the power series
$$ E_{\mu, \nu} (z) :=
    \sum_{k=0}^{\infty}\,
   \frac{z^k}{\Gamma(\mu \,k + \nu)}\,,
 \q z \in\CC\,. \eqno(B.6) $$
 It generalizes the classical Mittag-Leffler function
 to which it reduces for $\nu=1$.
 It is an entire  transcendental function of order $1/\Re \{\mu\} $
  on which the reader can inform himself  by again consulting 
  the  references  before outlined for the classical Mittag-Leffler function
\vsp
The function $E_{\mu, \nu}(-x)$  ($x\ge 0$)
 is completely monotonic if $0< \mu \le 1$ and  
  $\nu \ge \mu$.
 Again   this property is still valid
 if we consider the variable   $x = \lambda \,t^\mu$  where
 $\lambda$ is a positive constant. 
 In this case the asymptotic representations
 as $t\to 0^+$ and $t\to +\infty$ read
 $$ 
\null \!\!\! \!
 E_{\mu,\nu}(-\lambda t^\mu)
  \sim 
\cases{
{\ds \frac {1}{\Gamma(\nu)}  - \lambda \, \frac {t^\mu}{\Gamma(\nu+\mu)}},  
 & $ t \to 0^+,$ \cr\cr
{\ds \frac{1} {\lambda}\, \frac{t^{-\mu +\nu -1}}{\Gamma(\nu-\mu)} }\,, & $ t \to \infty.$ \cr
} 
\eqno(B.7)$$
We point out the Laplace transform pair, see \cite{Podlubny 99}, 
$$ \L\{t^{\nu-1}\, E_{\mu,\nu} (-\lambda t^\mu);s\} = 
\frac{s^{\mu -\nu} }{s^\mu + \lambda}\,,  \eqno(B.8)$$
with $\mu, \nu \in \RR^+$.
For $0< \mu = \nu \le 1$  this Laplace transform pair can be used
to derive 
 the noteworthy identity
 $$ t^{-(1-\mu )}
  E_{\mu ,\mu} \left(- \lambda\, t^{\mu }\right)
=  - \frac{1}{\lambda}\, \frac{d}{dt} E_\mu  \left (-\lambda\, t^{\mu }\right)\,,
\q 0<\mu \le 1\,.
\eqno(B.9)$$

\section*{Appendix C: The Exponential integral and \\ related functions}

The  exponential integral function, that we  denote by
$\EE_1(z)$, is defined as
$$ \EE_1(z)  
  =   \int_{z}^\infty \!\! \frac{\;\e^{\ds-t}}{ {\ds t}}\, dt
   = \int_1^\infty \frac{\e^{-zt}}{t} \, dt \,.
   \eqno(C.1)$$
We have used the letter $\EE$ instead of $E$ 
(commonly adopted  in the literature) 
in order to avoid confusion with  the Mittag-Leffler functions
that  play a more relevant role in fractional calculus.
This function exhibits a branch cut along the negative real semi-axis
and  admits the representation
$$ \EE_1(z)= -\gamma - \log z  - \sum _{n=1}^{\infty}
\frac{z^n}{n\, n!}\,, \; |\arg z| <\pi\,,
\eqno(C.2)$$
where $\gamma = 0.57721...$ is the so-called Euler-Mascheroni constant.
The power series  in the R.H.S. is absolutely convergent in all of $\CC$ 
and represents the entire function
called the {\it modified exponential integral}
$$ \Ein(z) :=  \int_0^z \frac{1-\e^{-\zeta}} {\zeta}\,d\zeta  =
- \sum _{n=1}^{\infty} \frac{z^n}{n\, n!}\,,
\eqno(C.3)$$
Thus, in view of (C.2) and (C.3), we write
$$ \EE_1(z) =   
   -\gamma  - \log z + \Ein (z)\,, \q |\hbox{arg}\, z| < \pi\,.
   \eqno(C.4)$$
	This relation is important for understanding the analytic properties
of the classical exponential integral function
in that it isolates the
multi-valued  part represented by the
logarithmic function from  the regular part
represented by the entire function  $\Ein(z)$. 
In $\RR^+$ the function $\Ein (x)$
 turns out to be a {\it Bernstein function},
which means that is positive, increasing,  with  the first derivative
{\it completely monotonic}.
 \vsp
The asymptotic behaviour as $z \to \infty$
of the  exponential integrals can be obtained
from the integral representation (C.1)
 noticing that
$$ \EE_1(z):= \int_{z}^\infty \!\! \frac{\;\e^{-t}}{ {\ds t}}\, dt
  =  \e^{\ds -z} \,
 \int_{0}^\infty \!\! \frac{\;\e^{-u}}{ {u+z}}\, du\,. \eqno(C.5)$$
In fact, by repeated partial integrations in the R.H.S.,  we get
for $| \hbox{arg} z | \le \pi -\delta\,,$
$$ \EE_1(z)  
   \sim \frac{\e^{\ds -z}}{ z}\,
   \sum_{n=0}^{\infty}\, (-1)^n \frac{n!}{ z^n}\,,
   \quad  z \to \infty\,.
 \eqno(C.6)
$$
We now report a number of relevant Laplace transform pairs
in which  logarithmic and exponential integral functions are involved. 
\vsp
Taking $t>0$, the basic Laplace transforms pairs are
$$ \L\{\log t; s\}  =- \frac{\gamma +\log s}{s}\,, \q  \Re s >0\,,
\eqno(C.7)$$
$$ \L\{\EE_1(t); s\}  = \frac{\log (s +1)}{s}\,, \q  \Re s >-1\,.
\eqno(C.8)$$
The proof of (C.7) and (C.8) is found, for example,
 in  the treatise  by Ghizzetti and Ossicini  \cite{Ghizzetti-Ossicini BOOK71},
 see Eqs. [4.6.15-16],  pp. 104-105.
 We then easily derive for $\Re s>0$
$$ \L\{\gamma + \log t; s\}  = - \frac{\log s}{s}\,, 
\eqno(C.9)$$
$$
\L\{\gamma +\log t + \EE_1(t); s\}  
=  \frac{\log (s +1)}{s} -\frac{\log s}{s}  
= \frac{\log (1/s +1)}{s}
 \,, 
 \eqno(C.10)$$
$$
\L\{\gamma +\log t + \e^{t}\,\EE_1(t); s\}  
= 
\frac{\log s}{s-1}- \frac{\log s}{s} 
 =\frac{\log s}{s(s-1)}
\,.
\eqno(C.11)$$
We  outline the different asymptotic behaviour of the three functions
$  \EE_1(t)$, $\Ein(t)$   
and $\gamma +\log t + \e^{t}\,\EE_1(t)\,,  $
for small argument ($ t \to 0^+$) and  large argument ($t \to +\infty$).
By using Eqs (C.2), (C.4) and (C.6), we have
$$  \EE_1(t)) \sim
\cases{
 \log (1/t) \,, & $ t \to 0^+\,,$ \cr\cr
 {\ds \e^{-t}/t}\,,  & $ t \to +\infty\,.$
}
\eqno(C.12) $$
$$  \Ein(t)  
\sim
\cases{
 t\,, & $t \to 0^+\,,$ \cr\cr
\log t   \,, &  $t \to +\infty\,.$
}
\eqno(C.13) $$
$$ 
\null \!\!\!
\gamma +\log t + \e^{t}\,\EE_1(t)
  \sim
\cases{
 t \, \log (1/t)  \,, & $t \to 0^+\,,$ \cr\cr
 \log t\,, &  $t \to +\infty\,.$
}
\eqno(C.14) $$
We note that all the above asymptotic representations can be obtained
 from the Laplace transforms of the corresponding functions by 
invoking the Tauberian theory for   {\it regularly varying    functions}  
(power functions multiplied by {\it slowly varying functions}\footnote{
{\bf Definition:} We call a (measurable) positive function $a(y)$,
 defined in a right neighbourhood of zero, {\it slowly varying at zero} if
$a(y)>0$ and
 $a(cy)/a(y) \to 1$ with $y \to 0$ for every $c>0$.
We call a (measurable) positive function $b(y)$,
 defined in a  neighbourhood of infinity, {\it slowly varying at infinity}
 if 
 $b(cy)/b(y) \to 1$ with $y \to \infty$ for every $c>0$.
 Examples: $(\log y)^{\gamma}$ with $\gamma \in \RR\,$ and
$\,\exp \,\left({\log y}/{\log\, \log y}\right)$.}), 
a topic   not so well known  which is
 adequately treated  in the treatise on Probability
 by Feller  \cite{Feller 71},  see Chapter XIII.5.
\vsp
We conclude this subsection pointing out
the Laplace transform pair
  $$ \L\{\nu (t, a); s\}  = \frac{1}{s^{a+1}\,\log s}\,, \q  \Re s >0\,.
\eqno(C.15)$$
where
$$\nu(t, a) := \int_0^\infty \, \frac {t^{a+\tau}} {\Gamma(a+ \tau +1)}\, d\tau\,,\q a > -1\,.
\eqno(C.16)$$
For details on  this transcendental function the reader is referred to the third volume of the Handbook 
of the Bateman Project \cite{Erdelyi HTF},
see  in Chapter XVIII (devoted to the Miscellaneous functions)
\S 18.3, pp.  217-224.


\begin{thebibliography}{99}

\bibitem{AnhLeonenko  JSP01}
Anh, V.V. and  Leonenko, N.N., 2001.
Spectral analysis of fractional kinetic equations with random data,
{\it J. Statistical Physics} {\bf 104}, 1349-1387 

\bibitem{AS 65}
   Abramowitz, M and Stegun, I.A., 1965.
  {\it  Handbook of Mathematical Functions},
 Dover, New York.




\bibitem{BagleyTorvik 00}
  Bagley, R.L. and  Torvik, P.J., 2000.
On the existence of the order domain and the solution of distributed
 order equations,  Part I, Part II,
{\it  International  Journal of  Applied Mathematics} {\bf 2},
 865-882, 965-987.







\bibitem{Caputo 67}
 Caputo, M., 1967.
  {Linear models of dissipation whose $Q$ is almost frequency
  independent,  Part II},
  {\it Geophys. J. R. Astr. Soc.} {\bf 13}, 529--539. 



\bibitem{Caputo 69}
  Caputo, M., 1969.
{\it Elasticit\`a e Dissipazione},
  Zanichelli, Bologna. (in Italian)


\bibitem{Caputo FERRARA95}
 Caputo, M,, 1995.
 Mean fractional-order derivatives differential equations and filters,
 {\it Annali della  Universit\`a di  Ferrara} (Sez VII, Sc. Mat.)
 {\bf 41}, 73-84.


 \bibitem{Caputo FCAA01}
 Caputo, M., 2001.
 Distributed order differential equations modelling
  dielectric induction and diffusion,
{\it Fractional Calculus and Applied Analysis} {\bf 4}, 421-442.  

\bibitem{CaputoMaina 71}
Caputo, M. and  Mainardi, F., 1971.
  Linear models of dissipation in  anelastic solids,
  {\it Rivista del  Nuovo Cimento\/} (Ser. II) {\bf 1}, 161--198.

\bibitem{ChechkinGorenfloSokolov PRE02}
Chechkin, A.V., Gorenflo, R.  and  Sokolov, I.M., 2002.
Retarding subdiffusion and accelerating superdiffusion
governed by distributed-order fractional diffusion equations,
{\it Physical Review E} {\bf  66}, 046129/1-6.

\bibitem{ChechkinGorenfloSokolovGonchar FCAA03}
Chechkin, A.V., Gorenflo, R., Sokolov, I.M. and Gonchar, V.Yu., 2002.
Distributed order time fractional diffusion equation,
{\it Fractional Calculus and Applied Analysis} {\bf 6}, 259-279.

\bibitem{ChechkinKlafterSokolov EUROPHYSICS03}
Chechkin, A.V., Klafter, J. and  Sokolov, I.M., 2003.
 Fractional Fokker-Planck  equation for ultraslow kinetics,
{\it Europhysics Lett.} {\bf 63}, 326-332.


 \bibitem{Diethelm-Ford FCAA01}
  Diethelm, K.  and  Ford, N.J., 2001.
 Numerical solution methods for distributed order
 differential equations,
 {\it Fractional Calculus and  Applied Analysis} {\bf  4}, 531-542.

 \bibitem{Diethelm-Luchko 04}
  Diethelm K. and  Luchko, Yu., 2004.
  Numerical solution of linear multi-term initial value problems
 of fractional order,
  {\it Computational  Methods of Numerical  Analysis} {\bf 6}, 243-263.


 \bibitem{Djrbashian 66}
  Djrbashian, M.M., 1966.
 {\it Integral Transforms and}
 {\it Representations of} {\it Functions in}
 {\it the Complex Plane},
   Nauka, Moscow. (in Russian).
  [There is also the transliteration as Dzherbashian]

  

\bibitem{Eidelman-Kochubei JDE04}
 Eidelman, S.D. and Kochubei, A.N., 2004.
 Cauchy problem for fractional diffusion equations,
{\it J. Differential Equations} {\bf 199}, 211-255.
  
 \bibitem{Erdelyi HTF}
   Erd\'elyi, A.,  Magnus, W.,  Oberhettinger, F.
 and Tricomi, F.G., 1955.
 {\it Higher Transcendental Functions},
 Vol. 3, McGraw-Hill, New York.

\bibitem{Feller 71}
   Feller, W., 1971.
 {\em An Introduction to Probability Theory and its Applications\/},
  Vol. 2, 2-nd Edition, Wiley, New York.

\bibitem{Gelfand-Shilov 64}
  Gel\`{}fand, I.M.  and Shilov, G.E., 1964.
 {\it Generalized Functions}, Volume I.
 Academic Press, New York and London. 

\bibitem{Ghizzetti-Ossicini BOOK71}
 Ghizzetti, A. and  Ossicini, A.,  1971.
{\it  Trasformate di Laplace e Calcolo Simbolico},
UTET, Torino. (in Italian)
  


 \bibitem{GoIsLu FCAA00}
 Gorenflo, R., Iskenderov, A. and Luchko, Yu., 2000.
 Mapping between solutions of fractional diffusion-wave equations,
 {\it Fractional Calculus and  Applied Analysis\/} {\bf  3}, 75--86.


\bibitem{GoLuMa 99}
Gorenflo, R.,  Luchko, Yu. and   Mainardi, F., 1999.
Analytical properties and applications of the Wright function,
{\it Fractional Calculus and  Applied Analysis} {\bf 2},  383-414.
\bibitem{GoLuMa 00}
Gorenflo, R., Luchko, Yu. and   Mainardi, F., 2000.
Wright functions as scale-invariant solutions of the diffusion-wave
 equation,
{\it Journal of Computational and Applied Mathematics\/}
     {\bf 118},   175-191.


\bibitem{GorMai CISM97}
  Gorenflo, R. and  Mainardi, F.,  1997.
 Fractional calculus:
  integral and differential equations of fractional order,
  in:  A. Carpinteri and F. Mai\-nardi (Editors),
  {\em Fractals and Fractional Calculus in Continuum Mechanics\/},
 Springer Verlag, Wien,   pp. 223--276.
  [Reprinted in  {\tt http://www.fracalmo.org}]

\bibitem{GorenfloMainardi CARRY05}
  Gorenflo, R. and  Mainardi, F., 2005.
   Simply and multiply scaled diffusion limits for continuous time random walks,
    {\it IOP (Institute of Physics) Journal of Physics: Conference Series}
   {\bf 7},  1-16.
   [S. Benkadda, X. Leoncini and  G. Zaslavsky (Editors):
   {Proceedings of the  International Workshop on Chaotic Transport and
   Complexity  in Fluids and Plasmas}
   Carry Le Rouet (France) 20-25 June 2004]






\bibitem{Hanyga TFD-PRS02}
 Hanyga, A., 2002.
Multi-dimensional solutions of time-fractional diffusion-wave equations,
 {\it Proc. R. Soc. London} {\bf A458}, 933-957.  
\bibitem{Hartley-Lorenzo SP03}
Hartley, T.T. and  Lorenzo, C.F., 2003.
Fractional-order system identification based on continuous
order-distributions,
{\it Signal Processing} {\bf 83}, 2287--2300.


\bibitem{Hilfer BOOK00}
 Hilfer, R. (Editor), 2000.
 {\it Applications of Fractional Calculus in Physics},
   World Scientific, Singapore.





\bibitem{Kilbas-et-al BOOK06}          
 Kilbas, A.A., Srivastava, H.M.  and  Trujillo, J.J., 2006.
 {\it Theory and Applications of Fractional Differential Equations},
 Elsevier, Amsterdam.
 
 
 \bibitem{Kochubei DE90}
 Kochubei, A.N., 1990.
 Fractional-order diffusion,    
{\it Differential Equations}  {\bf  26},  485--492.  
[English translation from the Russian Journal 
{\it Differenttsial'nye Uravneniya}]

 \bibitem{Langlands PhysA06}
 Langlands, T.A.M., 2006.
 Solution of a modified fractional diffusion equation,
{\it Physica A} {\bf 367}, 136-144.  
\bibitem{Lorenzo-Hartley NLD02}
Lorenzo, C.F. and  Hartley, T.T., 2002.
Variable order and distributed order fractional operators,
{\it Nonlinear Dynamics} {\bf 29}, 57-98.
\bibitem{Magin BOOK06}
Magin, R.L., 2006.
{\it Fractional Calculus in Bioengineering},
Begell House Pub., Connecticut, USA. 

\bibitem{Mainardi WASCOM93}
  Mainardi, F., 1993.
  On the initial value problem for the fractional diffusion-wave equation,
 in:  S. Rionero and T. Ruggeri (Editors),
 {\it Waves and Stability in Continuous Media},
  World Scientific, Singapore, pp. 246-251.


\bibitem{Mainardi CSF96}
   Mainardi, F., 1996.
  Fractional relaxation-oscillation and fractional
  diffusion-wave phenomena,
  {\it Chaos, Solitons and Fractals\/} {\bf 7}, 1461--1477.
\bibitem{Mainardi CISM97}
  Mainardi, F., 1997.
    Fractional calculus:
 some basic problems in continuum and statistical mechanics,
  in: A. Carpinteri and F. Mainardi (Editors),
  {\em Fractals and Fractional Calculus in Continuum Mechanics\/},
  Springer Verlag, Wien and New-York,  pp. 291--348.
[Reprinted in {\tt http://www.fracalmo.org}]

 \bibitem{MaiGor JCAM00}
 Mainardi, F. and  Gorenflo, R., 2000.
  On Mittag-Leffler type functions in fractional evolution processes,
  {\it J. Comput.  Appl. Math.} {\bf 118}, 283-299.

 \bibitem{MaLuPa FCAA01}
 Mainardi, F.,  Luchko, Yu. and  Pagnini, G., 2001.
     The fundamental solution of the space-time fractional diffusion
     equation,
   {\it Fractional Calculus and Applied Analysis}
 {\bf 4}, 153-192.
 [Reprinted in  {\tt http://www.fracalmo.org}]

  \bibitem{Mainardi-et-al SPRINGER-MME06}
 Mainardi, F., Mura, A.,  Pagnini, G. and  Gorenflo, R., 2007.
 Sub-diffusion  equations of fractional order
 and their fundamental solutions,
 in: K.Tas, J.A. Tenreiro Machado and D. Baleanu (Editors),
 {\it Mathematical Methods in Engineering},
 Springer Verlag, Dordrecht, pp. 23-55.
 [ISBN 978-1-4020-2 (HB)] 
 

 

 
 \bibitem{Mainardi-Pagnini AMC03}
Mainardi, F. and  Pagnini, G., 2003.
The Wright functions as solutions of the time-fractional diffusion equations,
{\it   Appl. Math. Comput.} {\bf 141},  51-62.
\bibitem{Mainardi-Pagnini JCAM06}
Mainardi, F. and  Pagnini, G., 2007.
The role of the Fox-Wright functions in fractional sub-diffusion of distributed order,
{\it   J. Comput. Appl. Math.} {\bf 207}, 245-257.

\bibitem{Mainardi-Pagnini-Gorenflo AMC06}
Mainardi, F., Pagnini, G. and Gorenflo, R., 2007.
Some aspects of fractional diffusion equations of single and distributed order,
{\it   Appl. Math. Comput.} {\bf 187}, 295-305. 


 \bibitem{MaPaSa JCAM05}
 Mainardi, F.,   Pagnini, G.  and  Saxena, R.K., 2005.
 Fox $H$ functions in fractional diffusion,
 {\it J. Comput. Appl. Math.}   {\bf  178},  321-331.                                  


 \bibitem{Metzler PhysA94}   
 Metzler, R.,  Gl\"ockle, W.G.   and   Nonnenmacher, T.F., 1994.
   Fractional model equation for anomalous diffusion,
   {\em Physica A\/} {\bf 211},  13--24.
\bibitem{Metzler-Klafter PhysRep00}
  Metzler, R. and  Klafter, J., 2000.
 The random walk's guide to anomalous diffusion: a fractional dynamics
 approach, {\it Physics Reports}  {\bf 339},   1-77.
\bibitem{Miller-Ross 93}
 Miller, K.S. and  Ross, B., 1993.
{\it An Introduction to the Fractional Calculus and Fractional Differential Equations},
Wiley, New York. 



 \bibitem{Naber FRACTALS04}
  Naber, M., 2004. 
 {Distributed order fractional subdiffusion},
 {\it Fractals} {\bf 12}, 23-32.  

 



\bibitem{Podlubny 99}
   Podlubny, I., 1999.
  {\it Fractional Differential Equations},
  Academic Press, San Diego.
  
  
  \bibitem{Saichev-Zaslavsky CHAOS97}
Saichev, A. and  Zaslavsky, G.M., 1997.
Fractional kinetic equations: solutions and applications,
 {\em Chaos\/} {\bf 7}, 753-764. 

 \bibitem{SKM 93}
  Samko, S.G.,  Kilbas, A.A. and  Marichev, O.I., 1993.
 {\it Fractional Integrals and Derivatives: Theory  and  Applications},
 Gordon and Breach, New York.
  
  
 
 

 \bibitem{SchneiderWyss 89}
 Schneider, W.R. and   Wyss, W., 1989.
  Fractional diffusion and wave equations,
  {\it Journal of  Mathematical Physics} {\bf 30}, 134-144.
\bibitem{SokolovChechkinKlafter POL04}
  Sokolov, I.M.,  Chechkin, A.V. and  Klafter, J., 2004.
 Distributed-order fractional kinetics,
{\it Acta Physica Polonica} {\bf 35}, 1323-1341.
\bibitem{SokolovKlafter CHAOS05}
 Sokolov, I.M. and  Klafter, J., 2005.
From diffusion to anomalous diffusion: a century after Einstein's
Brownian motion,
{\it Chaos} {\bf 15}, 026103-026109.
\bibitem{Umarov-Gorenflo ZAA05}
 Umarov, S. and Gorenflo, R., 2005.
 Cauchy and nonlocal multi-point problems for distributed order
pseudo-differential equations: Part one,
{\it Journal for Analysis and its Applications (ZAA)}  {\bf 24},  449-466.
\bibitem{West BOOK03}
  West, B.J., Bologna, M. and  Grigolini, P., 2003.
 {\it Physics of Fractal Operators}, Springer Verlag, New York.
\bibitem{Zaslavsky PhysRep02}
 G.M. Zaslavsky, G.M., 2002.
 Chaos,  fractional kinetics  and anomalous transport,
{\it Physics Reports}  {\bf 371},   461-580.
\bibitem{Zaslavsky BOOK05}
  Zaslavsky, G.M., 2005.
 {\it Hamiltonian Chaos and   Fractional Dynamics},
Oxford University Press, Oxford.

\end{thebibliography}
\end{document}